\documentclass[prl,floatfix,twocolumn,showpacs,preprintnumbers,amsmath,amssymb,superscriptaddress]{revtex4}
\usepackage{graphicx,float}

\newcommand{\udt}[3]{#1^{#2}_{\phantom{#2}#3}}
\newcommand{\dut}[3]{#1_{#2}^{\phantom{#2}#3}}
\newcommand{\dudt}[4]{#1_{#2\phantom{#3}#4}^{\phantom{#2}#3}}

\newcommand{\ket}[2][]{\mathinner{|#2\rangle}_{\hspace{-0.1em}#1}}

\begin{document}

\thispagestyle{empty}

\begin{center}
\title{\large{\bf The EPR correlation in Kerr-Newman spacetime}}
\date{29th November 2009}
\author{Jackson Levi Said}
\affiliation{Physics Department, University of Malta, Msida, Malta}
\author{Kristian Zarb Adami}
\affiliation{Physics Department, University of Malta, Msida, Malta}
\affiliation{Physics Department, University of Oxford, Oxford, OX1 3RH, United Kingdom}

\begin{abstract}
{The EPR correlation has become an integral part of quantum communications as has general relativity in classical communication theory, however when combined an apparent deterioration is
observed for spin states. We consider appropriate changes in directions of measurement to exploit full EPR entanglement for a pair of particles and show that it can be deduced only up to the
outer even horizon of a Kerr-Newman black hole, even in the case of freely falling observer.}
\end{abstract}


\maketitle

\end{center}

\section{I. Introduction}
For some of the founders of quantum mechanics one of the troubling parts was the spooky action-at-a-distance. Originally this was thought up by Einstein-Podolsky-Rosen (EPR) in an attempt to challenge certain aspects of quantum theory at the time. Contrary to its original design it is now the cornerstone of modern mainstream quantum physics, from cryptography to quantum computation, thus it is important to understand as many of the properties of quantum communications as possible. In particular it is of importance to fully understand the effect of spacetime curvature on EPR states. This is completely different from classical information transport. In this scenario the space between observer and emitter does not have an effect on the transmission which means that only local spacetime effects matter when making measurements on transmissions.
\newline
In this paper, we apply the Terashima and Ueda \cite{p1} approach to the spacetime background of a Kerr-Newman black hole. In general relativity the spin of a particle becomes deformed in all but the Minkowski spacetime. We present a method to extract the complete EPR correlation of two particles in a Bell state in Kerr-Newman geometry, ignoring helicity of infalling particles. These particles are defined locally and so suffer a precession of their spin component due an acceleration by an external force and the difference in the local inertial frame at different points about the given geometry. Taking these differences to arise from a continuous succession of local Lorentz transformations (LLT), the spin component can be calculated since it precesses in accord with the Wigner rotation. It is therefore not a trivial task to describe the motion of a particle using quantum mechanics near a Kerr-Newman black hole because the Poincar\'e group does not act intuitively in this region.
\newline
This paper is organized as follows, in Sec. II we derive the spin precession in the Kerr-Newman background for an observer at infinity. Then in Sec. III we consider the EPR correlation for a pair of fully entangled particles. In an attempt to remove the coordinate singularities from the derived angle we then calculate the spin precession for an infalling observer in Sec. IV. In Sec. V we discuss Bell's inequality for the observers at infinity and the infalling observer. Finally in Sec. VI we summarize our results.

\section{II. Kerr-Newman Distortion}
The most general vacuum solution of Einstein's field equations for black holes is the Kerr-Newman metric, any further complications requires one to consider hairy black holes. In this paper we take the Minkowski signature to be $\eta=\text{diag}(-,+,+,+)$ and use geometric units $(G=1=c)$. Latin letters are run over the four inertial labels $\it\left(\text{0,1,2,3}\right)$ and Greek letters over the four general coordinate labels. Also repeated indices are to be summed. Then the metric for the Kerr-Newman spacetime in {\it{Boyer-Lindquist\;coordinates}} $(t,r,\theta,\phi)$ for an observer at infinity is given by
\begin{widetext}
\begin{eqnarray}
ds^2\,=\,g_{\mu\nu}(x)dx^\mu dx^\nu\,=\,
    -\frac{\Delta}{\Sigma}(dt-a\,sin^2\theta\,d\phi)^2+
    \frac{\Sigma}{\Delta}dr^2+\Sigma d\theta^2+\frac{sin^2\theta}{\Sigma}(a\,dt-(r^2+a^2)\,d\phi)^2 \nonumber \\
\end{eqnarray}
\end{widetext}

where
\begin{align}
\Sigma\left(r\right)\,&=\,r^2+a^2\,cos^2\theta \nonumber\\
\Delta\left(r\right)\,&=\,r^2-2Mr+a^2+Q^2
\label{eq:init}
\end{align}
and Q, a and M the charge, angular momentum per unit mass and mass of the black hole respectively. For the most part the explicit statement of the dependence of Eq.(\ref{eq:init}) on $r$ will be assumed and so it will be suppressed for brevity unless otherwise stated. Together these three parameters form a family containing a set of all classical black holes. The coordinate system breaks down twice for this metric, firstly for the radial part when $\frac{1}{g_{rr}}=0$ then for the time part of the metric when $g_{tt}=0$,
\begin{equation}
r_{\pm}\,=\,M\pm\sqrt{M^2-a^2-Q^2}\nonumber\\
\label{hor}
\end{equation}
Where $r_{+(-)}$ is the outer (inner) event horizon. In order to relate local and global coordinates one must introduce a tetrad $\left(\text{or vierbein}\right)$ formalism, which we have chosen to be
\begin{align}
&\dut{e}{0}{\mu}(x)=\left(\frac{a^2+r^2}{\sqrt{ \Delta\Sigma}}, 0, 0, -\frac{a
   \sin (\theta )}{\sqrt{\Sigma}} \right)\nonumber\\
&\dut{e}{1}{\mu}(x)=\left(0, \sqrt{\frac{\Delta}{\Sigma}}, 0, 0\right) \nonumber\\
&\dut{e}{2}{\mu}(x)=\left( 0, 0, \frac{1}{\sqrt{\Sigma}}, 0 \right) \nonumber\\
&\dut{e}{3}{\mu}(x)=\left( -\frac{a}{\sqrt{\Delta\Sigma}}, 0, 0, \frac{1}{\sin(\theta)\sqrt{\Sigma}}\right)
\label{tetrad}
\end{align}

This describes a local inertial frame for a particle, in this case they are rotating with respect to an observer at infinity. In this respect Eq.(\ref{tetrad}) and its inverse are central to relating local and global coordinates. For example the momentum in a local inertial frame $p^{a}\left(x\right)$ is $\udt{e}{a}{\mu}\left(x\right)p^{\mu}\left(x\right)$ in relation to its global definition. It is imperative that properties can be related locally and globally on a manifold. Henceforth terms not shown are vanishing unless explicitly stated otherwise.

A straight-forward, but tedious calculation yields the components of the connection one-form \cite{p7}, $\dudt{\omega}{\mu}{a}{b}(x)=\udt{e}{a}{\nu}(x)\nabla_{\mu}\dut{e}{b}{\nu}(x)$, these are a spin connection. These are very involved equations in the Kerr-Newman spacetime. One notes however that the nonzero Schwarzschild symmetry is preserved and extended to some other pairs of elements. The following are the nonvanishing one-forms restricted to the equatorial plane,
\begin{align}
&\dudt{\omega}{t}{0}{1}(x)=\dudt{\omega}{t}{1}{0}(x)& &\dudt{\omega}{\theta}{1}{2}(x)=-\dudt{\omega}{\theta}{2}{1}(x)&\nonumber\\
&\dudt{\omega}{\varphi}{1}{3}(x)=-\dudt{\omega}{\varphi}{3}{1}(x)& &\dudt{\omega}{t}{1}{3}(x)=-\dudt{\omega}{t}{3}{1}(x)&\nonumber\\
&\dudt{\omega}{\varphi}{0}{1}(x)=\dudt{\omega}{\varphi}{1}{0}(x)& &\dudt{\omega}{r}{0}{3}(x)=\dudt{\omega}{r}{3}{0}(x)&\nonumber\\
&\dudt{\omega}{r}{0}{0}(x)=\dudt{\omega}{r}{3}{3}(x)& &&\nonumber\\
\end{align}

A particle in a circular orbit on the equatorial plane $(\theta=\pi/2)$ is now considered, with a radius r$(>r_+)$ and constant velocity. The four velocity of such a particle is given by
\begin{widetext}
\begin{align}
&u^t(x)=N^{-1}\cosh(\zeta)=r  \sqrt{\frac{\left(a^2+r^2\right)^2-a^2 \Delta }{a^2 \left(2 M r-Q^2\right)^2+\left(\Delta -a^2\right)
   \left(\left(a^2+r^2\right)^2-a^2 \Delta \right)}}\cosh (\zeta )\nonumber\\
&u^\varphi(x)=-N^{-1}N^{\varphi}\cosh(\zeta)+\frac{\sinh(\zeta)}{\sqrt{g_{\varphi\varphi}}}\nonumber\\
&\;\;\;\;\;\;\;\;\;\;=\frac{r}{\sqrt{\left(a^2+r^2\right)^2-a^2 \Delta }}\left[\frac{a  \left(2 M r-Q^2\right)\cosh (\zeta )}{\sqrt{a^2 \left(2 M
   r-Q^2\right)^2+\left(\Delta -a^2\right) \left(\left(a^2+r^2\right)^2-a^2 \Delta \right)}}+\sinh (\zeta )\right]
\label{BLu}
\end{align}
\end{widetext}

where $\zeta$ is the rapidity in the local inertial frame defined by
\begin{equation}
v=\tanh\left(\zeta\right)
\end{equation}
and
\begin{equation*}
N=\frac{1}{\sqrt{-g^{tt}}}
\end{equation*}
\begin{equation}
N^{\varphi}=\frac{g_{t\varphi}}{g_{\varphi\varphi}}
\end{equation}
subject to the constraint $u^{\nu}u_{\nu}=-1$ as stated in \cite{p5}, where $\textit{N}$ is the lapse function and $N^{\varphi}$ is the nonvanishing component of the shift vector field.
\newline
This is not however a geodesic, so an external force must be applied to the particle to counter the gravitational field. The acceleration, $a^{\mu}(x)=u^{\nu}(x)\nabla_{\nu}u^{\mu}(x)$, of such a force will be
\begin{widetext}
\begin{align}
a^r(x)=&-\frac{1}{r^3 \left(a^2 \left(r (2
    M+r)-Q^2\right)+r^4\right)^2}\Bigg[a \sqrt{\Delta } \sinh (2 \zeta ) \left(a^2 \left(M r-Q^2\right)+r^2 \left(3 M r-2 Q^2\right)\right) \big(a^2 \left(2 M r-Q^2+r^2\right)\nonumber\\
    &+r^4\big)+\frac{1}{2} r (M-r) \left(a^2 \left(2 M r-Q^2+r^2\right)+r^4\right)^2+\frac{1}{2} \cosh (2 \zeta
    ) \big(2 a^4 \left(Q^2-M r\right)+\nonumber\\
    &a^2 \left(r^2 \left(6 M^2-3 M r+r^2\right)+Q^2 r (3 r-7 M)+2 Q^4\right)+r^4 \left(r (r-3
    M)+2 Q^2\right)\big) \left(a^2 \left(2 M r-Q^2+r^2\right)+r^4\right)\Bigg]
\end{align}
\end{widetext}

The change in local inertial frame, $\udt{\chi}{a}{b}(x)=-u^{\nu}(x)\dudt{\omega}{\nu}{a}{b}(x)=u^{\nu}(x)\dut{e}{b}{\mu}(x)\nabla_{\nu}\udt{e}{a}{\mu}(x)$, between different points is now shown to be a boost along the 1-axis and a rotation about the 2-axis. In particular $\udt{\chi}{a}{b}(x)$ is a local quantity that relates the total change in local inertial frames along the path of the particle. Formally this quantity arises by the procedure that follows, in particular one first notes that a particle with four momentum $p^a\left(x\right)$ will suffer a change $\delta p^a\left(x\right)$ in its four momentum when moving between points $x^{\mu}$ and $x^{\mu}+u^{\mu}d\tau$ on the curved spacetime, where $d\tau$ is the infinitesimal proper time between events. The local change in four momentum corresponds to a global change by
\begin{align}
\delta p^a\left(x\right)&=\delta\left(p^{\mu}\left(x\right)\udt{e}{a}{\mu}\left(x\right)\right)\nonumber\\
&=\delta p^{\mu}\left(x\right)\udt{e}{a}{\mu}\left(x\right)+p^{\mu}\left(x\right)\delta\udt{e}{a}{\mu}\left(x\right)
\label{fou_mom}
\end{align}
The change in four momentum is given by taking the Fermi-Walker derivative
\begin{widetext}
\begin{align}
\delta u^{\mu}\left(x\right)&=-\left[u^{\nu}\left(x\right)u_{\nu}\left(x\right)a^{\mu}\left(x\right)-u_{\nu}\left(x\right)u^{\mu}\left(x\right)a^{\nu}\left(x\right)\right]d\tau\nonumber\\
&=-\left[u^{\mu}\left(x\right)a_{\nu}\left(x\right)-a^{\mu}\left(x\right)u_{\nu}\left(x\right)\right]u^{\nu}\left(x\right)d\tau
\end{align}
\end{widetext}
Thus
\begin{equation}
\delta p^{\mu}=-\frac{1}{m}\left[p^{\mu}\left(x\right)a_{\nu}\left(x\right)-a^{\mu}\left(x\right)p_{\nu}\left(x\right)\right]p^{\nu}\left(x\right)d\tau
\end{equation}
and the change in local inertial frame by
\begin{align}
\delta\udt{e}{a}{\mu}\left(x\right)&=u^{\nu}\left(x\right)d\tau\nabla_{\nu}\udt{e}{a}{\mu}\left(x\right)\nonumber\\
&=u^{\nu}\left(x\right)\dudt{\omega}{\nu}{a}{b}\left(x\right)\udt{e}{b}{\mu}\left(x\right)d\tau\nonumber\\
&=\udt{\chi}{a}{b}\left(x\right)\udt{e}{b}{\mu}\left(x\right)d\tau
\end{align}
where we have made use of the identity $\udt{e}{a}{\mu}\left(x\right)\dut{e}{b}{\mu}\left(x\right)=\udt{\delta}{a}{b}$ and indeed $\udt{\chi}{a}{b}=-u^{\nu}(x)\dudt{\omega}{\nu}{a}{b}\left(x\right)$ with $\dudt{\omega}{\nu}{a}{b}\left(x\right)=-\dut{e}{b}{\mu}\left(x\right)\nabla_{\nu}\udt{e}{a}{\mu}\left(x\right)=
\udt{e}{a}{\mu}\left(x\right)\nabla_{\nu}\dut{e}{b}{\mu}\left(x\right)$. Combing the above in to eq.(\ref{fou_mom}) gives
\begin{equation}
\delta p^a\left(x\right)=\udt{\lambda}{a}{b}\left(x\right)p^b\left(x\right)d\tau
\end{equation}
where
\begin{equation}
\udt{\lambda}{a}{b}\left(x\right)=-\frac{1}{m}\left[a^a\left(x\right)p_b\left(x\right)-p^a\left(x\right)a_b\left(x\right)\right]+\udt{\chi}{a}{b}\left(x\right)
\label{lambda_new}
\end{equation}
The explicit values for $\udt{\chi}{a}{b}\left(x\right)$ turn out to be
\begin{widetext}
\begin{align}
\udt{\chi}{0}{1}(x)&=\udt{\chi}{1}{0}(x)=\frac{\Delta}{a^4 r^2 \left(
   \Delta-a^2\right)+a^2 r^3 \left(4 M r \left( M-  r\right)+2r^3-Q^2 \left(a^2+2 M r-r^2\right)+\Delta  (2
   M-r)\right)+ r^6\left(\Delta-a^2 \right)}\nonumber\\
   &\Bigg[\left(2 a^2 r^2+r^4\right) \cosh (\zeta ) \left(Q^2-M r\right) \sqrt{\frac{\left(a^2+r^2\right)^2-a^2
   \Delta }{\left(\left(a^2+r^2\right)^2-a^2 \Delta \right) \left(r (r-2 M)+Q^2\right)+a^2 \left(Q^2-2 M r\right)^2}}-\nonumber\\
   &\frac{1}{\left(\left(a^2+r^2\right)^2-a^2 \Delta \right)^{3/2}}\bigg[a\Big(a^4 \left(-\left(r (r-3 M)+Q^2\right)\right)+a^2 \Big(Q^2 \left(a^2+3 M r+Q^2+r^2\right)+r \big(\Delta  (r-3 M)-\nonumber\\
   &r\left(6 M^2-3 M r+r^2\right)\big)-Q^4\Big)+r^4 \left(r (r-5 M)+3 Q^2\right)\Big) \nonumber\\
   &\left(\frac{a \cosh (\zeta )
   \left(\left(a^2+r^2\right)^2-a^2 \Delta \right) \left(2 M r-Q^2\right)}{\sqrt{\left(\left(a^2+r^2\right)^2-a^2 \Delta
   \right) \left(r (r-2 M)+Q^2\right)+a^2 \left(Q^2-2 M r\right)^2}}+\sinh (\zeta ) \left(a^2 \left(2 M
   r-Q^2+r^2\right)+r^4\right)\right)\bigg]\Bigg]
\end{align}
\end{widetext}

\begin{widetext}
\begin{align}
\udt{\chi}{1}{3}(x)&=-\udt{\chi}{3}{1}(x)\frac{\sqrt{\Delta }}{r^4}\Bigg[ \cosh (\zeta ) \left(\frac{a \left(2 M r-Q^2\right) \left(2 a^2 \left(Q^2-M
   r\right)+r^4\right)}{\sqrt{\left(a^2+r^2\right)^2-a^2 \delta }}+2 a \left(M r-Q^2\right)\right)\nonumber\\
   &\sqrt{\frac{\left(a^2+r^2\right)^2-a^2 \Delta }{\left(\left(a^2+r^2\right)^2-a^2 \delta \right) \left(-2 M
   r+Q^2+r^2\right)+a^2 \left(Q^2-2 M r\right)^2}}+\sinh (\zeta ) \left(a^2 \left(2 M r-Q^2+r^2\right)+r^4\right)\Bigg]
\end{align}
\end{widetext}

\begin{figure*}
\centerline{\includegraphics[width=15cm,height=5.3cm]{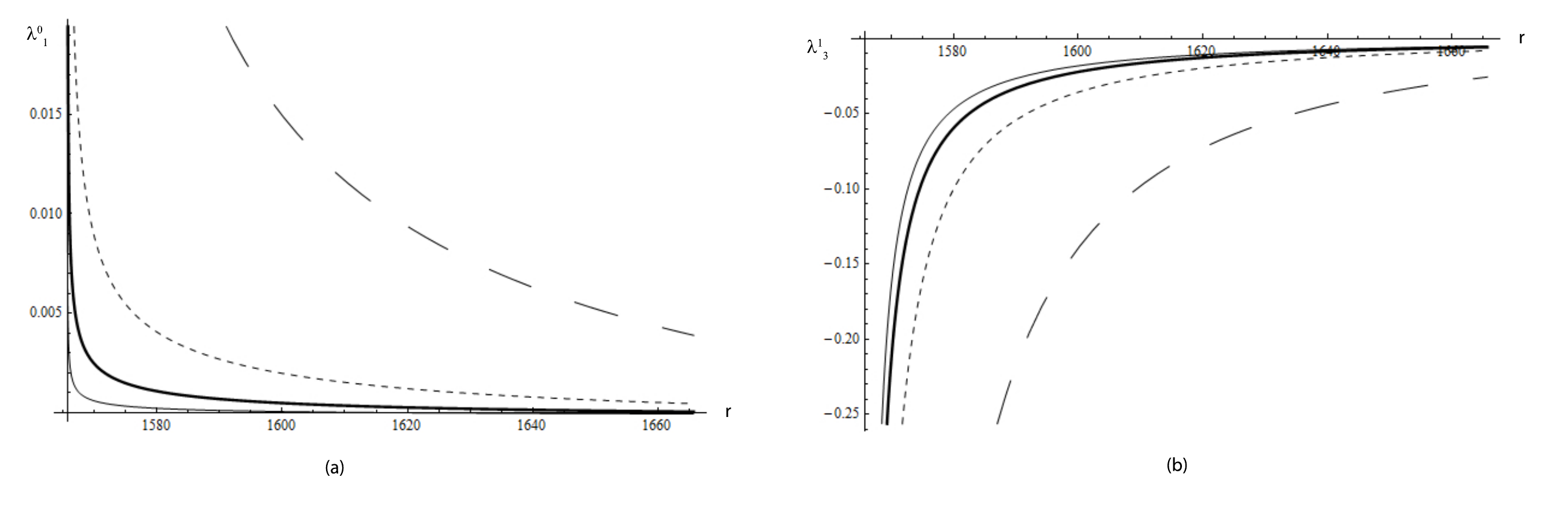}}
\caption{The infinitesimal LLT for a Kerr-Newman black hole with parameters $M=1000$, $a=0.8M$, $Q=0.2M$ and $v=0.3,0.5,0.7\text{ and }0.9$ for the black, thick, dashed and thick dashed lines respectively.}
\label{FigLam}
\end{figure*}

The LLT is defined by Eq.(\ref{lambda_new}), where $m$ is taken to be the mass of the particle species under consideration. This is an infinitesimal LLT since $\lambda_{ab}=-\lambda_{ba}$. The LLT then turns out to have only four nonvanishing terms which can be separated in to two symmetries,
\begin{align}
   \udt{\lambda}{0}{1}(x)=&\udt{\lambda}{1}{0}(x)\nonumber\\
   \udt{\lambda}{1}{3}(x)=&-\udt{\lambda}{3}{1}(x)
\end{align}
The explicit expressions are lengthy and so are represented graphically in Fig.(\ref{FigLam}) for some average parameters of a Kerr-Newman black hole.

\begin{figure*}
\centerline{\includegraphics[width=15cm,height=5.3cm]{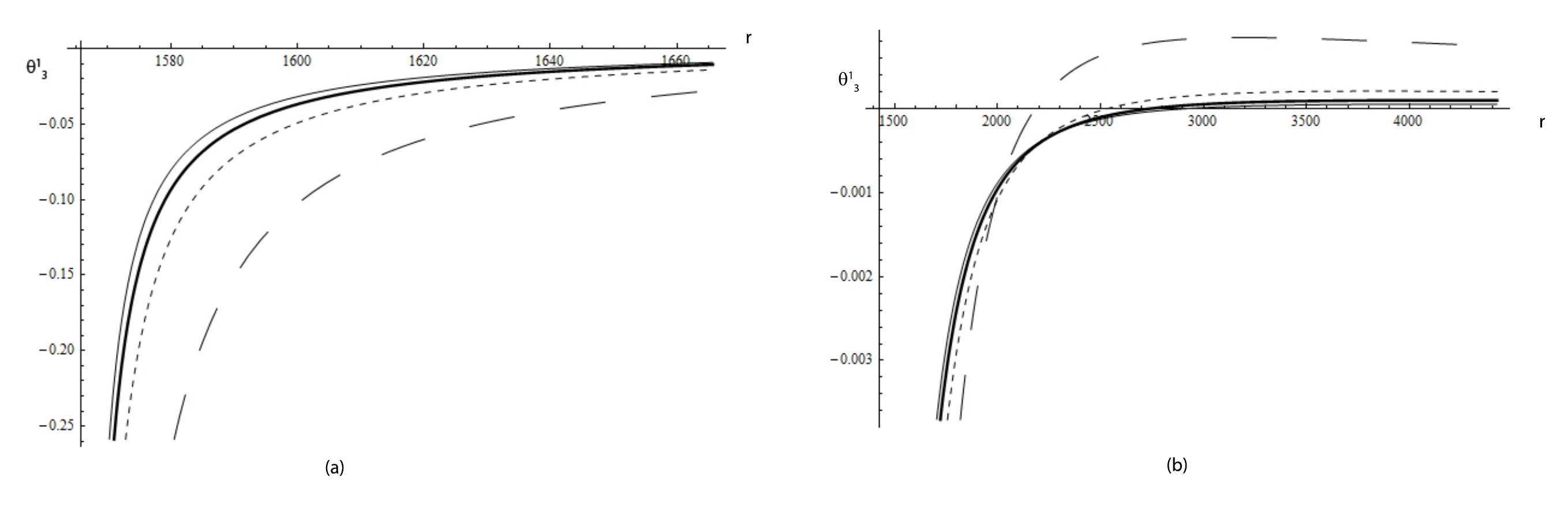}}
\caption{The rotation angle for a Kerr-Newman black hole with parameters $M=1000$, $a=0.8M$ and $Q=0.2M$ for the figure on the left and, $M=1000$, $a=0.9M$ and $Q=0.1M$ for the one on the right, and both with $v=0.3,0.5,0.7\text{ and }0.9$ for the black, thick, dashed and thick dashed lines respectively.}
\label{theta_fig}
\end{figure*}

This turns out to also be a boost along the 1-axis and a rotation about the 2-axis, as in Schwarzschild black hole case. With constant momentum $p^a\left(x\right)=\left(m\cosh\left(\zeta\right),0,0,m\sinh\left(\zeta\right)\right)$ pointing in the 3-axis, the change of the spin becomes a rotation as follows.
\vspace*{6pt}
\newline
After an infinitesimal proper time $d\tau$, the particle moves in the 3-axis by an amount, $\delta\phi=u^{\varphi}d\tau$. Over this the momentum in the local inertial frame transforms under the LLT, $\udt{\Lambda}{a}{b}\left(x\right)=\udt{\delta}{a}{b}+\udt{\lambda}{a}{b}\left(x\right)d\tau$, which corresponds to a unitary operator that acts on the state of the particle. This operator changes the spin, in particular it acts like the unitary matrix $U$ in
\begin{align}
U\left(\Lambda\left(x\right)\right)&\ket{p^a\left(x\right),\sigma;x}=\nonumber\\
&\displaystyle\sum_{\sigma'} D^{\left(1/2\right)}_{\sigma^{'}\sigma}\left(W\left(x\right)\right)\ket{\Lambda p^a\left(x\right),\sigma';x},
\label{spin_precession}
\end{align}
as in \cite{p8}, where $W\left(x\right)=W\left(\Lambda\left(x\right),\,p\left(x\right)\right)=L^{-1}\left(\Lambda p\right)\,\Lambda\,L\left(p\right)$ is a local Wigner rotation. It follows that $\udt{W}{a}{b}\left(x\right)=\udt{W}{a}{b}\left(\Lambda\left(x\right),\,p\left(x\right)\right)=\udt{\left[L^{-1}\left(\Lambda p\right)\,\Lambda\,L\left(p\right)\right]}{a}{b}$, with a standard Lorentz transform (LT) $\udt{L}{a}{b}\left(p\right)$ defined by,
\begin{align}
\udt{L}{0}{0}\left(p\right)=\gamma,\nonumber\\
\udt{L}{0}{i}\left(p\right)=\udt{L}{i}{0}\left(p\right)=\frac{p^i}{m},\nonumber\\
\udt{L}{i}{k}\left(p\right)=\delta_{ik}+\left(\gamma-1\right)\frac{p^ip^k}{|\overrightarrow{p}|^2},
\end{align}
where
\begin{equation}
\gamma=\frac{\sqrt{|\overrightarrow{p}|^2+m^2}}{m}\;\;\;\text{and}\;\;\;i,k\in\{1,2,3\}
\end{equation}
which turns out to be a rotation about the 2-axis though the angle $\udt{\vartheta}{1}{3}(x)=-\udt{\vartheta}{3}{1}(x)$, as will be shown in Eq.(\ref{Theta}). This is yet another tediously long equation, so we employ graphical techniques in Fig.(\ref{theta_fig}) to illustrate that a singularity is evident at the event horizon $r_+$. This is expected since an observer at infinity can not make measurements at and beyond an event horizon.

It is important here to note that $\udt{\vartheta}{a}{b}\left(x\right)\neq\udt{\lambda}{a}{b}\left(x\right)\neq\udt{\chi}{a}{b}\left(x\right)\neq\udt{\varphi}{a}{b}\left(x\right)$, where \begin{equation}
\udt{\varphi}{1}{3}\left(x\right)=-\udt{\varphi}{3}{1}\left(x\right)=u^{\varphi}\left(x\right)
\label{trivial}
\end{equation}
which is a trivial change in the local inertial frame. The former set of inequalities results from the boost part of $\udt{\lambda}{a}{b}\left(x\right)$, the acceleration and hence the force of the particle, and the curvature of the spacetime respectively.
\newline
Considering now the special case when $M,a,Q\rightarrow0$, the Minkowski spacetime is again recovered with

\begin{equation}
\label{chi31}
\udt{\chi}{1}{3}(x)=-\udt{\chi}{3}{1}(x)=\frac{\sinh(\zeta)}{r}
\end{equation}
and
\begin{equation}
\label{theta31}
\udt{\vartheta}{1}{3}(x)=-\udt{\vartheta}{3}{1}(x)=\frac{\cosh(\zeta)\sinh(\zeta)}{r}
\end{equation}
and most importantly for this situation, the Thomas precession of the spin , i.e. [Eq.(\ref{theta31})  -  Eq.(\ref{chi31})] for $v<<1$, remains
\begin{equation}
\left[\udt{\vartheta}{3}{1}(x)-\udt{\chi}{3}{1}(x)\right]d\tau\,\sim\,-\frac{va}{2 c^2}dt
\end{equation}
where $a\equiv|a^r(x)|=c^2\sinh^2(\frac{\zeta}{r})$. In this limit the change in the local inertial frame is just a rotation in the 2-axis through the angle $\udt{\chi}{1}{3}\left(x\right)$ and the change in spin is also a rotation about the 2-axis given by $\udt{\vartheta}{1}{3}\left(x\right)$. The difference between the latter two terms gives the spin precession in the low velocity limit per unit $dt=d\tau\cosh\left(\zeta\right)$

\begin{figure}[H]
\centering
\includegraphics[width=6cm, height=4.5cm]{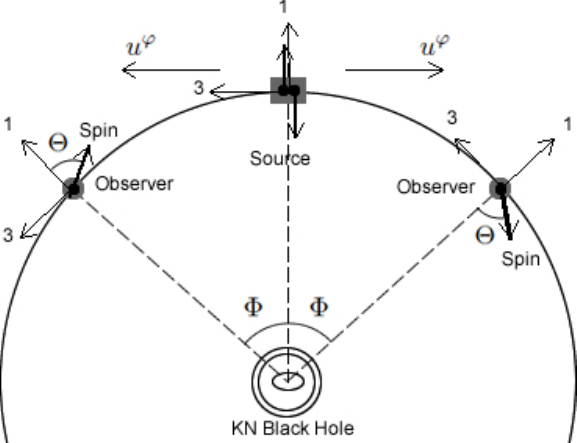}
\caption{\label{fig1} The EPR setup in the Kerr-Newman spacetime. Gray circles are the observers and the gray square is the EPR source.}
\end{figure}

\section{III. EPR Correlation}
We now move on to the actual gedanken experiment of this paper. Two observers are considered at azimuthal angles $\pm\Phi$ with $r>r_+$ and an EPR source generator at $\Phi=0$ as seen in Fig. (\ref{fig1}). The observers and EPR source are static in the local inertial frame Eq.(\ref{tetrad}), i.e. they are relatively static. The EPR source generates a pair of maximally entangled particles at $\Phi=0$ which are sent in opposite directions with constant four momentum $\udt{p}{a}{\pm}=\left(m\cosh\left(\zeta\right),0,0,\pm m\sinh\left(\zeta\right)\right)$. The pure state can be described by the Bell state,
\begin{equation}
\frac{1}{\sqrt{2}}\left[\ket{p^a_+,\uparrow;0}\ket{p^a_-,\downarrow;0}-\ket{p^a_+,\downarrow;0}\ket{p^a_-,\uparrow;0}\right],
\end{equation}
where for notational simplicity we put the $\phi-$coordinate in the state argument. After a proper time $\Phi/u^{\varphi}$ (since there is zero azimuthal acceleration) the particles reach their respective observers. An induced spin precession is observed because of the curved nature of the spacetime background, this is called the local Wigner rotation and in this case is given by
\begin{equation}
\udt{W}{a}{b}(\pm\phi,0)=\left({\begin{array}{cccc}
1 & 0 & 0 & 0\\
0 & \cos(\Theta) & 0 & \pm \sin(\Theta)\\
0 & 0 & 1 & 0\\
0 & \mp\sin(\Theta) & 0 & \cos(\Theta)\\
\end{array}}\right)
\label{unitary}
\end{equation}
where
\begin{equation}
\Theta=\Phi\frac{r}{\sinh(\zeta)}\udt{\vartheta}{1}{3}\\
\end{equation}
which is similar but not identical to the case for when $a=0=Q$ since the $\Theta$ is different. The induced precession arises from Eq.(\ref{spin_precession}), which moves in accord to the local Wigner rotation. For an infinitesimal LT
\begin{equation}
\udt{\Lambda}{a}{b}\left(x\right)=\udt{\delta}{a}{b}+\udt{\lambda}{a}{b}\left(x\right)d\tau
\end{equation}
The infinitesimal Wigner rotation corresponds to
\begin{equation}
\udt{W}{a}{b}\left(x\right)=\udt{\delta}{a}{b}+\udt{\vartheta}{a}{b}\left(x\right)d\tau,
\end{equation}
where $\udt{\vartheta}{0}{0}\left(x\right)=0=\udt{\vartheta}{0}{i}\left(x\right)=\udt{\vartheta}{i}{0}\left(x\right)$ and
\begin{equation}
\udt{\vartheta}{i}{k}\left(x\right)=\udt{\lambda}{i}{k}\left(x\right)+\frac{\udt{\lambda}{i}{0}p_k\left(x\right)-\lambda_{k0}\left(x\right)p^i\left(x\right)}{p^0\left(x\right)+m}
\label{Theta}
\end{equation}
In \cite{p1} it is shown that this is entirely equivalent to
\begin{align}
\udt{W}{a}{b}\left(x_f,x_i\right)=&\displaystyle\lim_{N\to\infty}\displaystyle\prod_{k=0}^{N} \left[\udt{\delta}{a}{b}+\udt{\vartheta}{a}{b}\left(x_{\left(k\right)}\right)\frac{h}{N}\right]\nonumber\\
=&T\exp\left[\displaystyle\int_{\tau_i}^{\tau_f}\udt{\vartheta}{a}{b}\left(x\left(\tau\right)\right)d\tau\right]
\end{align}
where $T$ is a time ordering operator and the exponent is the Taylor series exponent of the whole matrix. In this situation however the time ordering term is not needed since $\udt{\vartheta}{a}{b}\left(x\right)$ is constant during the motion of the particle (assuming that mass is not injected in to the system during this time).
\newline
The significant terms of the spin representation, i.e. the unity ones, can be condensed in to a Pauli representation as follows
\begin{align}
D^{\left(1/2\right)}_{\sigma^{'}\sigma}\left(W\left(x\right)\right)= I+\frac{i}{2}\Big[\vartheta_{23}&\left(x\right)\sigma_x+\vartheta_{31}\left(x\right)\sigma_y\nonumber\\
&+\vartheta_{12}\left(x\right)\sigma_z\Big]d\tau
\end{align}
where with the current parameters gives,
\begin{equation}
D^{\left(1/2\right)}_{\sigma^{'}\sigma}\left(W\left(\pm\Phi,0\right)\right)=\exp\left(\mp i\frac{\sigma_y}{2}\Theta\right)
\end{equation}
Thus when the particle pair reach their respective observers the state is described by
\begin{align}
\frac{1}{\sqrt{2}}\Big[\cos(\Theta)\big(&\ket{p^a_+,\uparrow;\Phi}\ket{p^a_-,\downarrow;-\Phi}\nonumber\\
&-\ket{p^a_+,\downarrow;\Phi}\ket{p^a_-,\uparrow;-\Phi}\big)\nonumber\\
+\sin(\Theta)\big(&\ket{p^a_+,\uparrow;\Phi}\ket{p^a_-,\uparrow;-\Phi}\nonumber\\
&+\ket{p^a_+,\downarrow;\Phi}\ket{p^a_-,\downarrow,-\Phi}\big)\Big]
\end{align}

as visible from Fig.(\ref{fig1}). This is where the entanglement appears to breakdown because of the presence of the spin triplet state in the state description. First of all however we remove the trivial rotation that rotates the local inertial frames by $\pm\Phi$ as in Eq.(\ref{trivial}). about the 2-axis at $\phi=\pm\Phi$. This is achieved through rotating the bases by $\mp\Phi$ about the 2-axis at $\phi=\pm\Phi$ respectively,
\begin{align}
\ket{p^a_\pm,\uparrow;\pm\Phi}'=&\cos(\frac{\Phi}{2})\ket{p^a_\pm,\uparrow;\pm\Phi}\nonumber\\
&\pm\sin(\frac{\Phi}{2})\ket{p^a_\pm,\downarrow;\pm\Phi},\\
\end{align}
\begin{align}
\ket{p^a_\pm,\downarrow;\pm\Phi}'=&\mp\sin(\frac{\Phi}{2})\ket{p^a_\pm,\uparrow;\pm\Phi}\nonumber\\
&+\cos(\frac{\Phi}{2})\ket{p^a_\pm,\downarrow;\pm\Phi},
\end{align}

As in \cite{p1},

\begin{align}
\frac{1}{\sqrt{2}}\Big[&\cos(\Delta)\big(\ket{p^a_+,\uparrow;\Phi}'\ket{p^a_-,\downarrow;-\Phi}'\nonumber\\
&-\ket{p^a_+,\downarrow;\Phi}'\ket{p^a_-,\uparrow;-\Phi}'\big)\nonumber\\
&+\sin(\Delta)\big(\ket{p^a_+,\uparrow;\Phi}'\ket{p^a_-,\uparrow;-\Phi}'\nonumber\\
&+\ket{p^a_+,\downarrow;\Phi}'\ket{p^a_-,\downarrow;-\Phi}'\big)\Big]
\end{align}
is found to describe the state, where
\begin{equation}
\Delta=\Theta-\Phi=\Phi\left[\frac{r}{\sinh(\zeta)}\udt{\vartheta}{1}{3}-1\right]
\end{equation}
Since the trivial rotation has been removed it is clear that a real deterioration of the perfect correlation between the spins is being observed, however only local unitary operations have been applied and the entanglement is invariant under unitary operations. Hence this must be an affect of the acceleration and gravity. If the pure state can be recovered then quantum computations may still be done while in the presence of a gravitational field. In particular the respective observers at $\phi=\pm\Phi$ must take measurements at an angle $\mp\Theta$ in their local inertial frames. Since $\udt{\vartheta}{a}{b}\left(x\right)\neq\udt{\chi}{a}{b}\left(x\right)$ a parallel transport would not reproduce this angle. Hence by transforming in the appropriate direction the full EPR correlation may still be recovered.
\newline
Now it was found that $\Delta$ is positive for $r\rightarrow\infty$ to a radius $r_0$ very close to the outer horizon $r_+$. As $r$ becomes smaller than $r_0$ and furthermore $r\downarrow r_+$, $\Delta\rightarrow-\infty$ and thus to extract the perfect EPR correlation each observer would require infinite accuracy in the measurement that even a small error would lead to a mixed state element.

\section{IV. The Infalling Observer}
We adopt the Doran \cite{p4} metric to remove the coordinate singularities of the metric. The observer can now fall through the apparent singularities of the Kerr-Newman spacetime observed by an observer at infinity. For this observer the line element is given by
\begin{widetext}
\begin{equation}
ds^2\,=\,-dT^2+\left[\frac{\Sigma}{\Omega}dR+b\,\frac{\Omega}{\Sigma}\left(dT-a\sin^2\left(\theta\right) d\phi\right)\right]^2+\Sigma^2d\theta^2+\Omega^2\sin^2\left(\theta\right) d\phi^2
\end{equation}
\end{widetext}
where
\begin{eqnarray}
&\Omega=\left(\text{R}^2+a^2\right)^{\frac{1}{2}}\nonumber\\
&b=\frac{\left(2MR-Q^2\right)^{\frac{1}{2}}}{\Omega}\nonumber\\
&\Sigma=\left(R^2+a^2\cos^2\left(\theta\right)\right)^{1/2}
\end{eqnarray}
and the time coordinate coincides with the proper time for the free fall observer. The vierbein is now chosen to be,
\begin{align}
&\dut{\tilde{e}}{0}{\mu}(x)=\left(1,0,0,0\right)\nonumber\\
&\dut{\tilde{e}}{1}{\mu}(x)=\left(b\,\frac{\Omega}{\Sigma},\frac{\Sigma}{\Omega},0,-a\,b\sin^2\left(\theta\right)\,\frac{\Omega}{\Sigma}\right) \nonumber\\
&\dut{\tilde{e}}{2}{\mu}(x)=\left(0,0,\Sigma,0\right) \nonumber\\
&\dut{\tilde{e}}{3}{\mu}(x)=\left(0,0,0,\Omega\,\sin\left(\theta\right)\right)
\label{Dvierbein}
\end{align}
In the $\left(t,r,\theta,\phi\right)$ coordinates the vierbein inherited the $r_+$ and $r_-$ coordinate singularities, the above $\left(T,R,\theta,\phi\right)$ also act as the metric does at those radii, and since the metric is singularity free there so is the vierbein. Now similarly to Eq.(\ref{BLu}) we take a four velocity of the form,
\begin{widetext}
\begin{align}
&\tilde{u}^T(x)=\tilde{N}^{-1}\cosh\left(\tilde{\zeta}\right)\nonumber\\
&\tilde{u}^\varphi(x)=\tilde{N}^{-1}\tilde{N}^{\varphi}\cosh\left(\tilde{\zeta}\right)+\frac{\sinh\left(\tilde{\zeta}\right)}{\sqrt{g_{\phi\phi}}}
\label{new_vel}
\end{align}
\end{widetext}
where $\tilde{\zeta}$ is the rapidity in the $\left(T,R,\theta,\varphi\right)$ local inertial frame. However constraining the two free variables $\tilde{N}^{-1}$ and $\tilde{N}^{\varphi}$ by the normalization equation $u^{\nu}u_{\nu}=-1$ yields
\begin{align}
&\tilde{N}=\frac{a \left(Q^2-2 M R\right)}{-2 a^2 M R+a^2 Q^2-a^2 R^2-R^4}\\
&\tilde{N}^{\varphi}=\pm\frac{\sqrt{-2 a^2 R M+a^2 Q^2-a^2 R^2-R^4}}{R\sqrt{\Delta\left(R\right)}}
\label{n_phi}
\end{align}
The four velocity thus emerges out of the normalization condition. Now Eq.(\ref{n_phi}) clearly will be singular on both horizons since they are defined as the solution to $\Delta=0$. In this way we find that even a freely falling observer can not extract the EPR correlation at and beyond the event horizon of a Kerr-Newman black hole. Thus no such observer can exist in this scenario which we attribute this to the strength of the frame dragging effects at the outer horizon since observers may be defined in the Schwarzschild black hole \cite{p1}. The result is a little surprising however it follows from the intrinsic nature of the Kerr-Newman black hole that no observer of this kind may be defined globally which is what would be required to make measurements on a bi-local property.
\newline
Furthermore following the same method as with the observer at infinity, the local Wigner rotation also turns out not to be finite on and beyond the out horizon which is a clear consequence of the absence of observability and {\it not} the lost of the EPR correlation. Hence it is the inability of an observer, so defined, to make measurements that restrict the regions where the EPR correlation can be extracted successfully.

\section{V. Maximum Violation of Bell's Inequality}
As in \cite{p1} we examined circularly moving particles in a local inertial frame. Similarly we measure the spin of one particle in the $(1,0,0)$-direction (component Q) or in the $(0,1,0)$-direction (component R) and the other particle in the $(-1,-1,0)$-direction (component S) or $(1,-1,0)$-direction (components T) in the local inertial frames $\phi=\Phi$ and $\phi=-\Phi$ respectively. In the Kerr-Newman geometry a decrease in the maximal violation of Bell's inequality is observed, in particular as
\begin{equation}
\langle QS\rangle+\langle RS\rangle+\langle RT\rangle-\langle QT\rangle=2\sqrt{2}\cos^2\left(\Theta\right)
\end{equation}
This however still includes the trivial rotation Eq.(\ref{trivial}) which could be causing the decrease in maximal violation of Bell's inequality. Rotating the components by $\mp\Phi$. The new spin components then become, $\left(\cos\left(\Phi\right),0,-\sin\left(\Phi\right)\right)$-direction (component $Q'$) or $\left(0,1,0\right)$-direction (component $R'$) for one particle and $\frac{1}{\sqrt{2}}\left(-\cos\left(\Phi\right),-1,-\sin\left(\Phi\right)\right)$-direction (component $S'$) or $\frac{1}{\sqrt{2}}\left(\cos\left(\Phi\right),-1,\sin\left(\Phi\right)\right)$-direction (component $T'$) for the other one. The violation of Bell's inequality still reduces from maximal as,
\begin{equation}
\langle Q'S'\rangle+\langle R'S'\rangle+\langle R'T'\rangle-\langle Q'T'\rangle=2\sqrt{2}\cos^2\left(\Delta\right)
\end{equation}
This is due to the gravitational field and accelerations involved. Taking in to account the general relativistic effect on spin measurements are thus take in the directions $\left(\cos\left(\Theta\right),0,-\sin\left(\Theta\right)\right)$-direction or $\left(0,1,0\right)$-direction for one particle and $\frac{1}{\sqrt{2}}\left(-\cos\left(\Theta\right),-1,-\sin\left(\Theta\right)\right)$-direction or $\frac{1}{\sqrt{2}}\left(\cos\left(\Theta\right),-1,\sin\left(\Theta\right)\right)$ for the other one in the same respective local inertial frames. However as the radius where the experiment takes place reduces to the outer horizon, a small error can build up which still causes maximal entanglement to be lost. This corresponds to a requirement of infinite accuracy in making measurements as the observer approaches infinitely close to the horizon.
\newline
Considering next the freely falling observer Eq.(\ref{Dvierbein}), the angle of precession is observed to become infinite on and beyond the outer event horizon $r_+$ since the observer looses the ability to take such measurements. Thus on the equatorial plane the EPR correlation can not be extracted and so infers a region where information may be lost for such an observer making measurements locally.

\section{VI. Conclusion}
We considered the EPR correlation of two accelerated particles in a Kerr-Newman background and found that the correlation apparently decreases as seen in the directions of flat spacetime as does the degree of violation of Bell's inequality. We derived the Wigner rotation and showed that maximal violation of Bell's inequality can be achieved through appropriate coordinate transformations of the local inertial frames. In this new inertial frame the EPR correlation can be extracted up to the outer event horizon which is to be expected for an observer at infinity.
\newline
However at the outer horizon $r_+$ and below, for both the observer at infinity and the free fall observer, the EPR correlation is unmeasurable. In particular the rotation angle for an observer approaches negative infinity on both counts and so the correlation will not be extracted once both particles have gone over the outer even horizon. This occurs because the flow of spacetime itself does not allow the experimental set up required to extract the EPR correlation. Hence due to frame dragging effects becoming so intense one cannot achieve the relatively static condition required for extraction and so no further measurements may be made on the particles in question which means that the information stored in their spin states will become irrecoverable. Despite the apparent loss of information as measured by such an observer, it is actually stored by the black hole up to the singularity where theory fails to predict what will  happen.

\section{Acknowledgments}
This work would not have been possible without the support and hospitality of Professor Steve Rawlings (Astrophysics, University of Oxford) and for this we thank him. Thanks are also due to Professor Steven Gull (Astrophysics, University of Cambridge) and the referees for important remarks and comments.

\end{document}